\documentclass[twocolumn,showpacs,preprintnumbers,amsmath,amssymb]{revtex4}

\usepackage{graphicx}
\usepackage{dcolumn}
\usepackage{bm}
\usepackage{setspace}

\usepackage[english]{babel}
\usepackage{amsmath}
\newcommand{\be}{\begin{equation}}
\newcommand{\ee}{\end{equation}}
\newcommand{\bea}{\begin{eqnarray}}
\newcommand{\eea}{\end{eqnarray}}

\begin{document}

\title{ Synchrony with Shunting Inhibition  }

\author{Sachin S. Talathi$^{1}$ }
\email{sachin.talathi@bme.ufl.edu}
\author{ Dong-Uk Hwang$^{1}$ }
\author{Abraham Miliotis$^{1}$ }
\author{Paul R. Carney$^{1}$}
\author{William L. Ditto$^{1}$ }


\affiliation{%
$^{1}$J Crayton Pruitt Department of Biomedical Engineering, University of Florida, Gainesville, FL  32611\\
}%
\date{\today}

\begin{abstract}

Spike time response curves (STRC's) are used to study the influence of synaptic stimuli on the firing times of a neuron oscillator without the assumption of weak coupling. They allow us to approximate the dynamics of synchronous state in networks of neurons through a discrete map, which can then be used to predict the stability of patterns of synchrony in the network. General theory for taking into account the contribution from higher order STRC terms, resulting from perturbations caused by synaptic stimuli lasting for more than one firing cycle of a neuron, in the approximation of the discrete map  is still lacking. Here we present a  general theory to account for higher order STRC corrections in the approximation of discrete map to determine the domain of 1:1 phase locked state in a network of two interacting neurons. We test the ability of this discrete map to predict the domain of 1:1 phase locked state in a network of two neurons interacting through a shunting synapse.

\end{abstract}

\maketitle

Synchronous rhythms in the brain, in particular gamma rhythms (20-80 Hz) are known to constitute a fundamental mechanism for cognitive tasks such as object recognition\cite{Mima_2001,Baudry_1999}, associative learning\cite{Gruber_2001,Gruber_2002} and processing of sensory information in the brain \cite{Engel_2001,Aoki_1999}. Early theoretical work by Wang and Rinzel \cite{Wang_1992}, demonstrated the ability for reciprocally connected interneurons to exhibit synchronous rhythms, thereby providing a mechanistic framework for synchrony exhibited by a purely inhibitory neuronal network; i.e., neurons interacting through a hyperpolarizing GABA$_{\text{A}}$ synapse ($E_{R}<V_{\text{rest}}$), where $E_{R}$ is the reversal potential of GABA$_{\text{A}}$ synapse and $V_{\text{rest}}$ is the neuronal resting-potential. Recently there has been a resurgent interest in shunting inhibition ($V_{\text{rest}}<E_{R}\le V_{\text{T}}$), where $V_{T}$ is the threshold for a neuron to generate an action potential, following the experimental findings that GABA$_{\text{A}}$ mediated inhibition in hippocampal slices of adult mammalian brains, is fast and shunting \cite{Bartos_2002,Vida_2006}. Through simulation studies on a realistic network model of coupled interneurons, the authors in \cite{Bartos_2007} have demonstrated that shunting enhances  robust gamma-band synchrony  in the network in the presence of moderately high heterogeneity. However due to the inherent complexity of the realistic network considered in \cite{Bartos_2007} , (a large network comprising of 200 interneurons, electrical coupling among neighboring pair of interneurons, the presence of synaptic propagation delay, fast synaptic decay time, strong synaptic conductance) it is difficult to elucidate the contribution of shunting to the enhancement of synchrony of the network in the gamma-band. Infact recent theoretical work \cite{Jeong_2007} suggests that shunting may enhance stable asynchronous states in the network of coupled interneurons. However the authors note that the observed difference in their analysis of synchrony induced through shunting synapse and those done by \cite{Bartos_2007} may lie in the different regimes of synaptic coupling (weak versus strong) and the lack of heterogeneity in the intrinsic firing rates of the coupled neurons. 

Here, we investigate whether the analytical framework of spike time response curves (STRC's) can be used to predict the synchronous state of 1:1 phase locking between two coupled neurons interacting through a strong shunting synapse in the presence of heterogeneity. We begin by demonstrating that synaptic stimuli through shunting inhibition to a neuron periodically firing in the gamma frequency band persists for 3 consecutive firing cycles as opposed to the case for a neuron receiving strong synaptic stimuli through a hyperpolarizing synapse. We then develop an analytical framework to approximate the dynamics of 1:1 synchronous state in a network of two neurons coupled via a strong shunting synapse through a discrete map by taking into account the higher order STRC contributions. We show that in the limit of zero higher order STRC contribution's the discrete map reduces to the well-known approximation for synchrony between two neurons coupled through a hyperpolarizing inhibitory synapse \cite{Ermentrout_1996,Acker_2003,Talathi_2008}.
  
Each neuron in the network considered here is modeled based on a single compartment conductance based model for a fast spiking interneuron developed by  \cite{Wang_Buzsaki_1996}. The dynamical equation for the model neuron is given by \begin{eqnarray}
C\frac{dV(t)}{dt}&=&I_{DC}+I_{S}(t)+g_{Na}m^{3}_{\infty}h(t)(E_{Na}-V(t))  \nonumber \\
&+&g_{K}n^{4}(t)(E_K-V(t))+g_{L}(E_L-V(t)) 
\label{eqn1}
\end{eqnarray}
where V(t) is the membrane potential, $I_{DC}$ is the constant input DC-current that determines the intrinsic spiking period $T_{0}$, $E_{r}$ and $g_{r}$ represent the reversal potential and conductance of ion channels with parameters obtained from \cite{Wang_Buzsaki_1996}. The inactivation variable $h(t)$ and the activation variable $m_{\infty}$ for sodium channel and the activation variable $n(t)$ for the potassium channel satisfy the first order kinetic equations as described in \cite{Wang_Buzsaki_1996}. $I_{S}(t)=g_{s}S(t)(E_{R}-V(t))$ is the synaptic current with $E_{R}$ (mV)  being the reversal potential of the synapse and $g_{s}$ ($\frac{mS}{cm^{2}}$) being  the strength of synaptic coupling. $S(t)$, gives the fraction of bound receptors and satisfy the following first order kinetic equation as described in \cite{Abarbanel_2003,Talathi_2008}.
\begin{figure}
\includegraphics[scale=.4]{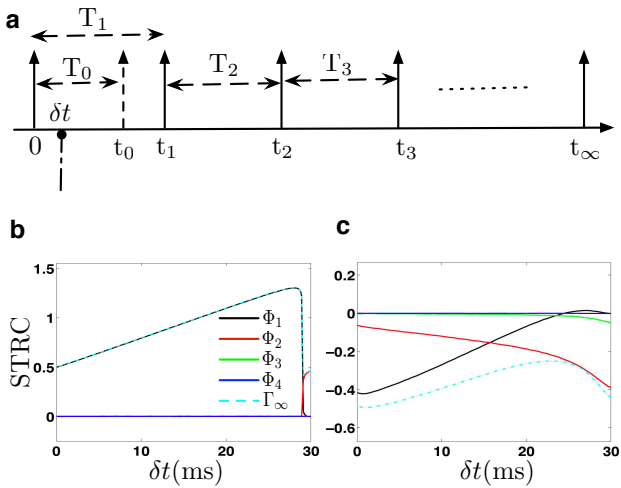}
\caption{\label{Fig1} (a) Schematic diagram demonstrating the effect of perturbation received by a spiking 
neuron at time t. The cycle containing the perturbation defines the first order STRC and the 
subsequent cycles define the higher order STRC terms. (b) The STRCÕs for hyperpolarizing synaptic input with $E_{R}=-80$ mV. (c) The STRCÕs for shunting synaptic input with $E_{R}=-55$ mV.  The synaptic parameters are $\tau_{R} = 0.1$ ms, $\tau_{D} = 8$ ms, $g_{s} = 0.15$ mS/cm$^{2}$ . 
The intrinsic period of Þring for the neuron is $T_{0} = 31$ ms and $V_{\text{rest}}=-65$ mV.}
\end{figure}
\begin{figure}
\includegraphics[scale=.4]{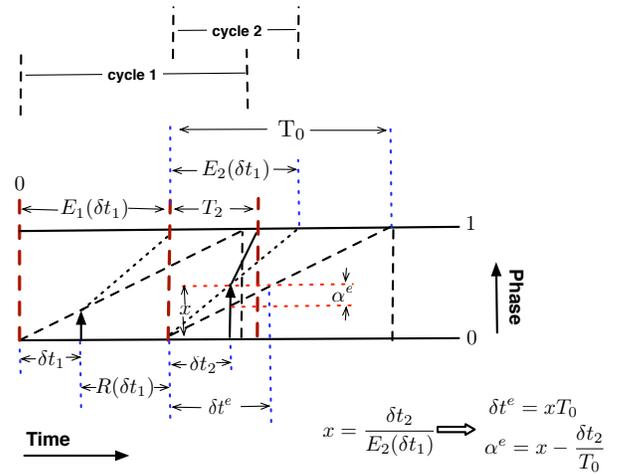}
\caption{\label{Fig2} Schematic diagram to demonstrate the re-normalization and re-scaling procedure to determine the length of second cycle $T_{2}$. Shown in red dashed lines are the effective spike times after the neuron receives two consecutive synaptic perturbations. Shown in black dotted lines is the chage in the firing cycle caused by synaptic input in the first firing cycle. Shown in black dashed line is the unperturbed firing cycle for the neuron.}
\end{figure}
\begin{figure}
\includegraphics[scale=.325]{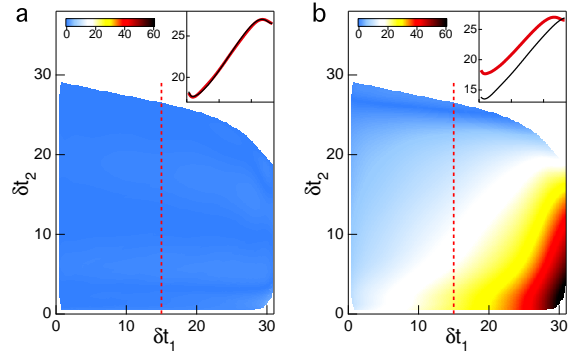}
\caption{\label{Fig3}In (a) and (b) we show the color coded percent error $E=100\left|\frac{T^{N}_{2}-T^{P}_{2}}{T^{N}_{2}}\right|$ between the predicted value of $T_{2}$, given by $T^{P}_{2}$ through STRC's in equations 2 and 3 respectively and the actual value of $T_{2}$ given by $T^{N}_{2}$ as determined by solving equation 1, numerically for neuron receiving two consecutive synaptic stimuli in successive firing. Inset shows the plot for the variation in $T_{2}$ ($T^{N}_{2}$ shown in red and $T^{P}_{2}$ shown in black) as a function of $\delta t_{2}$ for  $\delta t_{1}=15$ ms  }
\end{figure}

As a measure of the influence of synaptic stimuli on the firing times of a neuron, we define the spike time response curves (STRC's)  $\Phi_{j}(t,\tau_{R},\tau_{D},g,E_{R},T_{0})=\frac{T_{j}-T_{0}}{T_{0}}$ \cite{Oprisan_2004, Maran_2008, Talathi_2008}, where $T_{0}$ is the intrinsic period of spiking, $T_{j}$ represents the length of the $j^{th}$ spiking cycle from the cycle $j=1$ in which the neuron receives synaptic stimuli at time $0<t<T_{0}$. The synaptic parameters are: $\tau_{R}$: the synapse rise time, $\tau_{D}$: the synaptic decay time, $g$: the synaptic strength and $E_{R}$: the reversal potential of the synapse. In general the synaptic input need not be weak. The STRC's are obtained numerically using the direct method \cite{Acker_2003} as shown in the schematic diagram in Figure \ref{Fig1}a. The neuron firing regularly with period $T_{0}$, is perturbed through an inhibitory synapse at time $t$ after the neuron has fired a spike at reference time zero. The spiking time for neuron is considered to be the time when the membrane voltage V, crosses a threshold (set to 0 mV in all the calculations presented here). As a result of this perturbation, the neuron fires the next spike at time $t_{1}$, representing the first cycle after perturbation of length $T_{1}$, which is different from $T_{0}$, the time at which the neuron would have fired a spike in absence of any perturbation through inhibitory synapse. Depending on the properties of the synapse, i.e., synapse parameters: $\tau_{R}$, $\tau_{D}$ and $E_{R}$;  through which the neuron receives the perturbing input, the length of subsequent cycles might change. For example, as can be seen from Figure \ref{Fig1}b, for GABA$_{\text{A}}$ mediated inhibition through synapse that is hyperpolarizing, the first order STRC is non-zero for all perturbation times $0<\delta t<T_{0}$; the second order STRC is non-zero for $\delta t \rightarrow T_{0}$ and all higher order STRC terms are zero. However for GABA$_{\text{A}}$ mediated inhibition through synapse that is shunting (Figure \ref{Fig1}c) the effect of perturbation lasts for the first three cycles including the perturbing cycle, i.e., both the first order and second order STRC's is non-zero for $0<\delta t<T_{0}$ and the third order STRC is non-zero for $\delta t \rightarrow T_{0}$. Higher order STRC terms for shunting synapse are present because, shunting tends to depolarize the neuron thereby reducing the effective time for the occurrence of the next spike following synaptic stimuli. The asymptotic STRC $\Gamma_{\infty}$, gives the long term effect of perturbation received by a spiking neuron and is given through a linear sum of the first $m$ non-zero STRC's, i.e., $\Gamma_{\infty}\overset{\text{def}}{=}\sum^{m}_{j=1}\Phi_{j}$. For hyperpolarizing synapse shown in Figure \ref{Fig1}b, $m=2$ and for shunting synapse shown in Figure \ref{Fig1}c, $m=3$. For all further calculations, unless otherwise mentioned, we will suppress the dependence of STRC on synaptic parameter's and the intrinsic period of the neuron $T_{0}$. We further define the following two functions derived from STRC's: $R(\delta t)=T_{0}\left(1+\Phi_{1}(\delta t)\right)-\delta t$ and $E_{j}(\delta t)=T_{0}(1+\Phi_{j}(\delta t))$.

We will now use STRC's to  develop a general framework to determine an approximate discrete map for 1:1 synchrony between two coupled interneurons interacting through a shunting synapse. We will begin by considering the simple case of a periodically firing neuron that receives synaptic stimuli through a shunting synapse in each of its two successive firing cycles at times $\delta t_{1}$ and $\delta t_{2}$ as shown in Figure \ref{Fig2}. Our goal is to determine the length of the second cycle $T_{2}$ in this situation.
In the presence of a single perturbation at $\delta t_{1}$ in the cycle 1, following from the definition of STRC's we have $T_{2}=E_{2}(\delta t_{1})=T_{0}(1+\Phi_{2}(\delta t_{1}))$. Similarly in the presence of a single perturbation in cycle 2 at time $\delta t_{2}$, again following from the definition of STRC's we have $T_{2}=\delta t_{2}+R(\delta t_{2})=T_{0}(1+\Phi_{1}(\delta t_{2}))$. In writing this equation we note that the default period of second cycle in the absence of the single perturbation at time $\delta t_{2}$  was $T_{0}$. However if the synaptic input at $\delta t_{2}$ is followed by a synaptic perturbation in the previous cycle at $\delta t_{1}$, the default length of the second cycle is no longer $T_{0}$. Therefore, in order to correctly determine the length of second cycle in this case, we have to discount for the change in the default length of second cycle caused by synaptic input at time $\delta t_{1}$. We do so by re-normalizing the synaptic perturbation time in the second cycle to $\delta t^{e}_{2}=\delta t_{2}\frac{T_{0}}{E_{2}(\delta t_{1})}$ and re-scaling the effective phase of perturbation by $\alpha^{e}_{2}=\frac{\delta t_{2}}{E_{2}(\delta t_{1})}-\frac{\delta t_{2}}{T_{0}}$, as shown in Figure \ref{Fig2}a. The re-normalization of the perturbation time and the re-scaling of the perturbing phase, allows us to discount for the effect of synaptic input at time $\delta t_{1}$ on the length of cycle 2. The length of cycle 2 now becomes $T_{2}=\delta t_{2}+R(\delta t^{e}_{2}).(1-\alpha^{e}_{2})$. In terms of STRC's we have,
\bea
T_{2}&=&\delta t_{2}+\left(T_{0}\left(1+\Phi_{1}\left(\frac{\delta t_{2}}{1+\Phi_{2}(\delta t_{1})}\right)\right)-\frac{\delta t_{2}}{1+\Phi_{2}(\delta t_{1})}\right) \nonumber \\
&\times&\left(1-\frac{\delta t_{2}}{T_{0}(1+\Phi_{2}(\delta t_{1}))}+\frac{\delta t_{2}}{T_{0}}\right)
\eea
For a neuron receiving synaptic stimuli through a hyperpolarizing synapse, we have $\Phi_{2}(x)\approx 0$ (see Figure \ref{Fig1}b). In this case, equation 1 reduces to 
\bea
T_{2}\approx T_{0}(1+\Phi_{1}(\delta t_{2})+\Phi_{2}(\delta t_{1}))
\eea
The approximation in the form of equation 3, has been used in \cite{Oprisan_2001} to determine the effect of second order STRC component on stability of 1:1 synchronous state in a ring of pulse coupled oscillators; in \cite{Oprisan_2004} to determine phase resetting and phase locking in a hybrid circuit of one model neuron and one biological neuron and also recently in \cite{Maran_2008} to predict 1:1 and 2:2 synchrony in mutually coupled network of interneurons with synapse that is hyperpolarizing. In Figure \ref{Fig3}a, we plot the percent error $E=100\left|\frac{T^{N}_{2}-T^{P}_{2}}{T^{N}_{2}}\right|$
 between the predicted value for the length of second cycle: $T^{P}_{2}$, determined from equation 2, and the actual value: $T^{N}_{2}$, determined by numerically solving the ODE in equation 1, for a neuron receiving two consecutive synaptic stimuli at times $\delta t_{1}$ and $\delta t_{2}$ through a shunting synapse with following parameters: $\tau_{R}=0.1$ ms, $\tau_{D}=8$ ms, and $E_{R}=-55$ mV. In inset of Figure \ref{Fig3}a, we plot $T^{N}_{2}$ and $T^{P}_{2}$ for the specific case of $\delta t_{1}= 10$ ms. In Figure \ref{Fig3}b, we show similar results from predicted value of $T_{2}$ estimated using equation 3. We see that while equation 2 is correctly able to predict the length of second cycle,  equation 3  fails to capture the effect of second order STRC contribution in determining $T_{2}$. We would like to emphasize that the expression for $T_{2}$ is only dependent on STRC's estimated for a given synapse type without any explicit assumption on the strength of synaptic input to the neuron and is valid both in the regime of weak and strong coupling and for slow and fast synaptic dynamics.
 
We can now generalize our approach of re-normalization and re-scaling to consider the situation when the neuron receives 3 synaptic stimuli in 3 consecutive firing cycles at time $\delta t_{1}$ in cycle 1, at time $\delta t_{2}$ in cycle 2 and at time $\delta t_{3}$ in cycle 3. This is an important case to consider in order to determine an approximate discrete map for 1:1 synchrony between two neurons interacting through a shunting synapse; since we know from Figure \ref{Fig1}b that the effect of synaptic stimulus through a shunting synapse last for 3 consecutive firing cycles of the neuron. Following equation 2, the length of 3rd cycle in the presence of 3 consecutive synaptic stimuli can be written as: $T_{3}=\delta t_{3}+R(\delta t^{e}_{3}).(1-\alpha^{e}_{3})$; where $\delta t^{e}_{3}=\delta t_{3}\frac{T_{0}}{\tilde{E}_{3}(\delta t_{1},\delta t_{2})}$ and $\alpha^{e}_{3}\approx\frac{\delta t_{3}}{\tilde{E}_{3}(\delta t_{1},\delta t_{2})}-\frac{\delta t_{3}E_{2}(\delta t_{1})}{(T_{0})^{2}}$ where $\tilde{E}_{3}(\delta t_{1},\delta t_{2})=E_{2}(\delta t^{2}_{e})$, represents the length of third cycle in the presence of two consecutive synaptic stimuli at times $\delta t_{1}$ and $\delta t_{2}$.   Note that although $\delta t^{e}_{3}$ and $\alpha^{e}_{3}$ are dependent on timing of the occurrence of synaptic input in the first cycle, we have neglected the explicit contribution of $\Phi_{3}(\delta t_{1})$ in the effective re-scaling term $\alpha^{3}_{e}$ for determining $T_{3}$. This is a reasonable approximation since $\Phi_{3}(\delta t)\approx 0$ for shunting synapse (see Figure \ref{Fig1}b) resulting in $E_{3}(\delta t_{1})\approx T_{0}$.  Details on our  estimate of the approximation for $\alpha^{e}_{3}$ is presented elsewhere \cite{Talathi_2009}. 

\begin{figure}
\includegraphics[scale=.4]{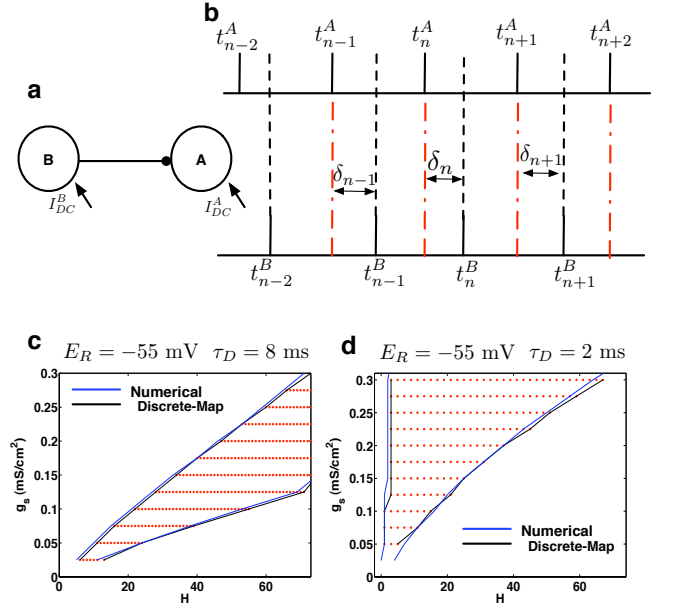}
\caption{\label{Fig4} (a) Schematic diagram of the network considered. (b) Schematic diagram representing spike timing for neurons A and B when they are phase locked in 1:1 synchrony. In (c) and (d) we show domain of 1:1 synchrony estimated through STRC's from the discrete map in equation 5 (shown in black) and those obtained through numerical simulations of the network (shown in blue) for network of two interneurons coupled with shunting synapse with parameters $tau_{R}=0.1$ ms, $E_{R}=-55$mV, $tau_{D}=8$ ms, and $tau_{D}=2$ ms respectively.}
\end{figure}

We are now in the position to derive the approximate discrete map for 1:1 synchrony between neurons A and B firing with intrinsic period $T^{A}_{0}\ne T^{B}_{0}$ and coupled through a shunting synapse (see Figure \ref{Fig4}a). The heterogeneity in the intrinsic firing rates of the two coupled neurons is quantified through $H=100\cdot\frac{I^{B}_{DC}-I^{A}_{DC}}{I^{A}_{DC}}$, where $I^{A}_{DC}$ and $I^{B}_{DC}$ are constant DC currents driving neurons A and B.
 From Figure \ref{Fig4}b, when the two neurons are locked in stable 1:1 synchrony we have for neuron A, $t^{A}_{n+1}=t^{A}_{n}+\delta_{n}+R\left(\delta_{n}\frac{T^{A}_{0}}{\tilde{E}_{3}(\delta_{n-2},\delta_{n-1})}\right).(1-\alpha^{e}_{3})$, where $t^{X}_{n}$ is the time of $n^{th}$ for neuron X=\{A,B\} and $\delta_{n}=\delta t^{B}_{n}-\delta t^{A}_{n}$. Since neuron B, does not receive any external perturbation, we have for neuron B, $t^{B}_{n+1}=t^{B}_{n}+T^{B}_{0}$. The discrete map for evolution of $\delta_{n}$ can then be obtained as: 
\bea
\delta_{n+1}&=&T^{B}_{0}-R\left(\delta_{n}\frac{T^{A}_{0}}{\tilde{E}_{3}(\delta_{n-2},\delta_{n-1})}\right).(1-\alpha^{e}_{3})
\eea
The steady state solution to above equation can be obtained by solving for the fixed point $\delta^{*}$ of the discrete map defined by $\delta_{n+1}=\delta_{n}=\delta_{n-1}=\delta_{n-2}=\delta^{*}$. We then obtain $F(\delta^{*})=T^{B}_{0}$, where $F(\delta^{*})$ is given by
\bea
F(\delta^{*})&\approx&\delta^{*}+R\left(\frac{\delta^{*}T^{A}_{0}}{\tilde{E}_{3}(\delta^{*},\delta^{*})}\right)\nonumber \\
&\times& \left(1-\frac{\delta^{*}}{\tilde{E}_{3}(\delta^{*},\delta^{*})}+\frac{\delta^{*}{E}_{2}(\delta^{*})}{(T^{A}_{0})^{2}}\right) 
\eea
In the limit of $\Phi_{2}\approx 0$, $F(\delta^{*})=T^{A}_{0}(1+\Phi_{1}(\delta^{*})+\Phi_{2}(\delta^{*}))$ corresponding to the well-known equation for the solution to the fixed point of discrete map for synchrony between two neurons coupled through a hyperpolarizing synapse \cite{Talathi_2008}. Stability of the fixed point $\delta^{*}$ representing the solution to equation 5 requires $0<\left|\frac{\partial F(\delta_{n},\delta_{n-1},\delta_{n-2})}{\partial (\delta_{n},\delta_{n-1},\delta_{n-2})}\right|_{\delta_{n}=\delta_{n-1}=\delta_{n-2}=\delta^{*}}<2$. This stable fixed point then corresponds to the synchronous 1:1 locked state for the two coupled interneurons. 
In order to determine whether equation 4 can predict 1:1 phase locked states for the two neuron network interacting through a shunting synapse, we consider the specific case of neurons A and B coupled through a shunting synapse with parameters: $E_{R}=-55$ mV, $\tau_{R}=0.1$ ms, and $\tau_{D}=8$ ms. Neuron A receives fixed dc current $I^{A}_{DC}$, such that it is firing with intrinsic period of $T^{A}_{0}=31$ ms. We solve equation 5, for different values of $H$, thereby modulating $T^{B}_{0}$, to determine the set of values for $g_{s}$, which will result in stable fixed point solution for equation 5. The solution is obtained by estimating STRC's for each value of $g_{s}$ and then determining whether there is a fixed point solution to equation 5. In Figure \ref{Fig4}c, we present the results of this calculation. For a given value of $H$, the curve in black gives the lower and upper bounds on the strength of coupling for shunting synapse $g_{s}$, for which a unique stable solution to equation 5 exists. For example with H=50, the range of values for $g_{s}$ for which a unique stable solution exists for equation 5 is $0.09<g_{s}<0.21$. This region of 1:1 synchronous locking is analogous to the classic Arnold tongue \cite{Kurths_2001, Talathi_2008}, obtained for synchrony between two coupled nonlinear oscillators. Arnold tongue provides a two dimensional visualization of  this dependence, as a bounded domain of region in the heterogeneity (H)-coupling strength (g) plane, where 1:1 synchrony between the two oscillators exist. 
In Figure \ref{Fig4}b, the general feature of the Arnold tongue is represented as the region bounded by two black curves obtained through STRC by solving for fixed point of equation 5. In Figure \ref{Fig4}b, shown in blue is a similar bound on the range of heterogeneity leading to synchronous oscillations between the two coupled neurons, obtained by numerically solving equation 1 for the evolution of the dynamics of the coupled neuron network. This curve is obtained by fixing the firing period of neuron A, $T^{A}_{0}$ and varying the firing period of neuron B, by changing $I^{B}_{DC}$ and determining the strength of synaptic coupling $g_{s}$ that results in $\frac{T^{B}_{0}}{\left<T^{A}\right>}\approx 1$. As can be seen from Figure \ref{Fig4}b, the results match to those obtained through STRC calculations for fixed point of equation 5. In Figure \ref{Fig4}d, we present similar calculation for the two neurons coupled through a fast shunting synapse with parameters: $E_{R}=-55$ mV, $\tau_{D}=2$ ms, and $\tau_{R}=0.1$ ms.\\ \\
We note that the Arnold tongue showed in Figure \ref{Fig4}c and Figure \ref{Fig4}d are skewed to the right; i.e., $H>0$, which suggests that shunting inhibition tends to promote 1:1 synchrony at higher frequencies of the driver neuron. Recently \cite{Talathi_2008} have demonstrated the mechanism for phase locked state of 1:1 synchrony to exhibit identical synchrony, which is essential for the generation of synchronous oscillations in a larger network of neurons.  It is therefore very likely then that through mechanism of spike timing dependent plasticity \cite{Talathi_2008}, the two neurons may lock in identical synchrony at frequencies in the gamma range. This may result in the generation of gamma oscillations in larger network of interneurons interacting through shunting inhibition as has been demonstrated through simulation studies of realistic networks of neurons \cite{Bartos_2007}. In conclusion, we have provided a general  framework to study synchrony between neurons coupled through a shunting inhibition using discrete maps derived from spike time response curves.

\bibliographystyle{apsrev}
\bibliography{ArnoldInhib}

\end{document}